\documentclass{article}

\usepackage{PRIMEarxiv}

\usepackage[utf8]{inputenc} 
\usepackage[T1]{fontenc}    
\usepackage{hyperref}       
\usepackage{url}            
\usepackage{booktabs}       
\usepackage{amsfonts}       
\usepackage{nicefrac}       
\usepackage{microtype}      
\usepackage{lipsum}
\usepackage{fancyhdr}       
\usepackage{graphicx}       
\graphicspath{{media/}}     

\usepackage{amsmath}
\usepackage{float}
\usepackage{subcaption}
\usepackage{adjustbox}

\usepackage{draftwatermark}
\SetWatermarkLightness{ 0.9 }
\SetWatermarkText{PREPRINT}
\SetWatermarkScale{1}

\title{Convolutional Neural Networks for Segmentation of Malignant Pleural Mesothelioma: Analysis of Probability Map Thresholds (CALGB 30901, Alliance)

}

\author{
  Mena Shenouda \\
  Department of Radiology \\
  The University of Chicago \\
  Chicago, IL, USA\\
  \texttt{shenmena@uchicago.edu} \\
   \And
   Eyjólfur Gudmundsson \\
   Icelandic Radiation Safety Office \\
   Reykjavik, Iceland \\
   \texttt{eyjolfur88@gmail.com}
   \And
   Feng Li \\
   Department of Radiology \\
   The University of Chicago \\
  Chicago, IL, USA\\
  \texttt{feng.li@uchospitals.edu}
  \And
   Christopher M. Straus \\
   Department of Radiology \\
   The University of Chicago \\
  Chicago, IL, USA\\
  \texttt{cstraus@uchicago.edu}
  \And
   Hedy L. Kindler \\
   Department of Radiology \\
   The University of Chicago \\
  Chicago, IL, USA\\
  \texttt{hkindler@medicine.bsd.uchicago.edu}  
   \And
  Arkadiusz Z. Dudek \\
  Metro Minnesota Community Oncology Research Consortium\\
  St. Louis Park, MN, USA \\
  \texttt{adudek021@outlook.com} \\
   \And
   Thomas Stinchcombe \\
   Duke Cancer Institute \\
   Duke University \\ 
   Durham, NC, USA \\
   \texttt{thomas.stinchcombe@duke.edu}
  \And
  Xiaofei Wang \\
  Alliance Statistics and Data Management Center \\
  Duke University \\
  Durham, NC, USA \\
  \texttt{xiaofei.wang@duke.edu}
    \And
   Adam Starkey \\
   Department of Radiology \\
   The University of Chicago \\
  Chicago, IL, USA\\
  \texttt{starkey@uchicago.edu}
    \And
   Samuel G. Armato III \\
   Department of Radiology \\
   The University of Chicago \\
  Chicago, IL, USA\\
  \texttt{s-armato@uchicago.edu}
}

\begin{document}
\maketitle



\begin{abstract}
Malignant pleural mesothelioma (MPM) is the most common form of malignant mesothelioma, with exposure to asbestos being the primary cause of the disease. To assess response to treatment, tumor measurements are acquired and evaluated based on a patient's longitudinal computed tomography (CT) scans. Tumor volume, however, is the more accurate metric for assessing tumor burden and response. Automated segmentation methods using deep learning can be employed to acquire volume, which otherwise is a tedious task performed manually. The deep learning-based tumor volume and contours can then be compared with a standard reference to assess the robustness of the automated segmentations. The purpose of this study was to evaluate the impact of probability map threshold on MPM tumor delineations generated using a convolutional neural network (CNN). Eighty-eight CT scans from 21 MPM patients were segmented by a VGG16/U-Net CNN. A radiologist modified the contours generated at a 0.5 probability threshold. Percent difference of tumor volume and overlap using the Dice Similarity Coefficient (DSC) were compared between the standard reference provided by the radiologist and CNN outputs for thresholds ranging from 0.001 to 0.9. CNN annotations consistently yielded smaller tumor volumes than radiologist contours. Reducing the probability threshold from 0.5 to 0.1 decreased the absolute percent volume difference, on average, from 43.96\% to 24.18\%. Median and mean DSC ranged from 0.58 to 0.60, with a peak at a threshold of 0.5; no distinct threshold was found for percent volume difference. The CNN exhibited deficiencies with specific disease presentations, such as severe pleural effusion or disease in the pleural fissure. No single output threshold in the CNN probability maps was optimal for both tumor volume and DSC. This study emphasized the importance of considering both figures of merit when evaluating deep learning-based tumor segmentations across probability thresholds. This work underscores the need to simultaneously assess tumor volume and spatial overlap when evaluating CNN performance. While automated segmentations may yield comparable tumor volumes to that of the reference standard, the spatial region delineated by the CNN at a specific threshold is equally important.
\end{abstract}


\section{Introduction} \label{sec:Intro}
Malignant pleural mesothelioma (MPM) is an aggressive form of cancer present in the pleural lining of the lungs. It is often the result of exposure to asbestos and has a poor prognosis \cite{Gerwen}. Computed tomography (CT) is the most common imaging modality used to stage and assess patients with MPM. The current standard to evaluate tumor response to therapy is the modified Response Evaluation Criteria in Solid Tumors (mRECIST). Using this protocol, observers make up to six measurements of “tumor thickness residing perpendicular to the chest wall or mediastinum” as presented on a CT scan \cite{Byrne}. This protocol has been recently updated as mRECIST 1.1 \cite{Eisenhauer}.

In contrast to the linear measurements of mRECIST, manual volumetric analysis conducted by radiologists can be used to better estimate tumor burden. Tumor volume has displayed strong predictive power in patient assessment in terms of overall and progression-free survival \cite{Pass, Murphy}; however, acquisition of manual tumor volume is too time-consuming and burdensome to be systematically used in routine clinical care. The difficulties of tumor volume calculations can be mitigated using artificial intelligence and machine learning, specifically convolutional neural networks (CNNs). Previous studies have successfully implemented CNNs in MPM segmentation, which is a crucial step in automating tumor volume measurements as the laborious part is performed automatically, and simple pixel counting remains for the volume calculation \cite{EG1, EG2}. 

For a CNN to properly segment MPM tumor on CT sections, the network must be trained and validated using a labeled set of images. Due to the fundamental statistical nature of machine learning, the outputs of a CNN are probabilities. For instance, in identifying the location of tumor on a CT scan, the CNN assigns each pixel a probability of being tumor. Therefore, for each CT section input to the network, a probability map is generated that displays the likelihood of tumor in a pixel-wise fashion. In practice, a threshold is set to binarize these maps so that any pixel with probability equal to or greater than 0.5, for example, is set to 1 (``tumor''), and all other pixels are set to 0 (``not tumor''). Given that the 0.5 probability of might not generate the most accurate segmentation for a disease as complicated as MPM tumor, the purpose of the present study was (1) to better understand the probability values returned by the CNN and (2) to investigate whether different probability thresholds could improve pixel-wise class labeling in this complicated tumor. The tumor segmentations obtained using different thresholds were evaluated against the reference standard using the Dice Similarity Coefficient (DSC) and the percent difference of volume as the figures of merit. Therefore, we investigated the impact that varying thresholds would have on the predicted tumor segmentation/probability maps, given the inherent complexities of this tumor.

Overall, the choice of threshold may have considerable impact on the final tumor segmentation. In terms of MPM tumor volumetry, a lower threshold applied to the probability maps would result in a larger volume as more pixels are now considered tumor; however, this may result in an increase of pixels erroneously labeled as tumor, thus negatively impacting the CNN accuracy of tumor segmentation. In contrast, increasing the threshold may produce a segmentation that is too restrictive, substantially decreasing the tumor volume calculated. Furthermore, the change in threshold alters the overlap of tumor identified by the CNN with the ground truth (as determined by a trained observer). It is important to note that the purpose of this work was not to assess an observer's performance when given initial computerized outlines (as our group has previously shown the significant effect of initial outlines on observer performance \cite{Sensakovic}) but rather to demonstrate the impact a choice of probability threshold has on defining tumor. Therefore, this study investigated the impact of probability thresholds on tumor volume and the overlap of tumor contours by applying a broad range of thresholds, recording the volumes and the DSC and studying the resulting trends.

\section{Methods} \label{sect:methods}
\subsection{Patient population}
The patient cohort was compiled from a previous study performed by the Cancer and Leukemia Group B (CALGB 30901) \cite{Dudek}. CALGB is now part of the Alliance for Clinical Trials in Oncology. The CALGB 30901 study evaluated 49 patients with confirmed unresectable epithelioid, sarcomatoid, or mixed-type MPM, and patients were without disease progression after 4 to 6 cycles on first-line therapy with pemetrexed and cisplatin or carboplatin. Patients were randomly assigned to either the treatment arm (continued therapy with pemetrexed alone) or the observation arm. The patients underwent CT scans at baseline and then every 6 weeks for the first 6 months. Each participant signed an IRB-approved, protocol-specific informed consent document in accordance with federal and institutional guidelines.

\subsection{CT image acquisition and tumor contours}
The present study was retrospectively conducted on 88 baseline and follow-up CT scans of 21 patients from the CALGB 30901 study. There was an average of 106 sections per scan (range: 39-418 sections) and 3.7-mm thickness (range: 1.25-5 mm). The scans were acquired using 21 different scanner models.

Tumor contours were generated automatically using a CNN previously published \cite{EG2} without any additional training or validation. The CNN employed the U-Net deep CNN (2D) architecture. Specifically, the deep CNN architecture consisted of a downsampling and upsampling path. For the downsampling, a Visual Geometry Group 16 (VGG16) model was pre-trained on the ImageNet database using scale-jittering \cite{UNet, VeryDeep}. Layers of the downsampling path were initialized using the weights acquired from the VGG16 training scheme. A 2$\times$2 max pooling operation with stride 2 was applied to the feature maps at each step of the downsampling path. A dropout layer of probability 0.5 was used to prevent overfitting. During the upsampling path, a 2D operation using nearest-neighbor interpolation was applied to the feature maps. The network output a segmentation mask the same size as the input image size (e.g., 512$\times$512 pixels). A rectified linear unit (ReLU) activation function was applied after the convolutional layers, except for the last layer, which used a sigmoid activation function that returned pixel-wise probabilities on the range [0,1] for the segmentation task, i.e., whether a pixel contains tumor. We applied a threshold of 0.5 to the output of the sigmoid layer during the validation step so that any pixel with a probability 0.5 or greater was labeled ``tumor.'' During its training phase, the network minimized the binary cross-entropy, which was averaged over all pixels of each predicted segmentation and the provided reference standard. Adam, an algorithm for first-order gradient-based stochastic optimization, was used to optimize the network during training using a learning rate of 10$^{-5}$. 

The VGG16/U-Net deep CNN architecture was previously trained on 126 MPM patients, some presenting with pleural effusion \cite{EG2}. In this earlier study, the CNN was tested on 34 patients, who all presented with both tumor and pleural effusion, and on 43 patients, some of whom presented with both tumor and pleural effusion and some only with tumor; the median DSC and median average Hausdorff distance (AHD) for that method were 0.690 and 5.1 mm, respectively, as previously reported \cite{EG1}. For the present study, no additional training or validation was performed before evaluating the performance of the trained CNN on the external CALGB dataset.  

A research radiologist [F.L.] presented with the CNN contours (acquired using a predetermined probability threshold value of 0.5) was able to modify/redraw the contours using in-house software to provide the reference standard. Due to the time-consuming nature of adjusting the contours, however, the radiologist was presented with sections separated by approximately 5 mm. This process resulted in an average of 51 reviewed sections per scan (range: 32-70 sections). Section comparisons and tumor volumes were performed only on contours of sections that the radiologist reviewed. 

\subsection{Tumor volume and Dice similarity coefficient}
\subsubsection{Tumor volume calculation}
Tumor volume was defined as:
\begin{multline}
    \text{Volume [mm}^3] = \sum \text{Number of pixels within a contour} \\ \times \text{pixel dimension}^2 \, \text{[mm}^2] \times \text{inter-section distance [mm]}
    \label{eq:volume}
\end{multline}
where the summation is over all sections containing a contour. The number of pixels within a contour was equal to the number of nonzero pixels within the binary mask created after applying a threshold to the probability maps generated by the CNN. Pixel dimension (in units of $mm^2$) was acquired from the DICOM header. Inter-section distance corresponded to the difference in table position between two sections on which the radiologist provided reference contours. All tumor volumes computations were performed using MATLAB (Mathworks Inc., Natick, Massachusetts).

\subsubsection{Dice similarity coefficient}
Another metric used to compare the CNN tumor contours generated at the different thresholds with the radiologist’s reference standard was the Dice similarity coefficient (DSC) \cite{Dice}:
\begin{equation}
    DSC(A,B) = \frac{2|A \cap B|}{|A|+|B|}
\end{equation}
Where $|A|$ and $|B|$ are the number of elements within a set, e.g., pixels within a contour. The DSC is a measure of overlap between segmentations: a comparison between contours generated by the CNN and the reference provided by the radiologist \cite{Metrics}. The DSC was calculated for each individual CT section (using MATLAB’s ``dice'' function), and a final DSC was calculated per patient after averaging the DSC values across all sections.

\subsection{Statistical methods}
Comparison of the absolute percent difference of volumes and comparison of DSC values across thresholds were first checked for normality using the one-sample Kolmogorov-Smirnov test \cite{KV_test}. Since the null hypothesis was rejected, none of the data was deemed to come from a standard normal distribution. Therefore, the Wilcoxon signed rank test was used to compare DSC or absolute percent volume differences computed between the reference standard contours and contours generated across a range of CNN probability map threshold values. Statistical significance was considered at $p=0.05$; the p-value was not adjusted to account for multiple comparison due to the small sample size in this hypothesis-generating study. Data collection was conducted by the Alliance Statistics and Data Management Center. Data quality was ensured by review of data by the Alliance Statistics and Data Management Center and by the study chairperson following Alliance policies.

\section{Results}
\begin{figure}[H]
    \centering
    \setkeys{Gin}{width=\linewidth}
    \begin{subfigure}{.4009\textwidth}
        \includegraphics[width=\linewidth]{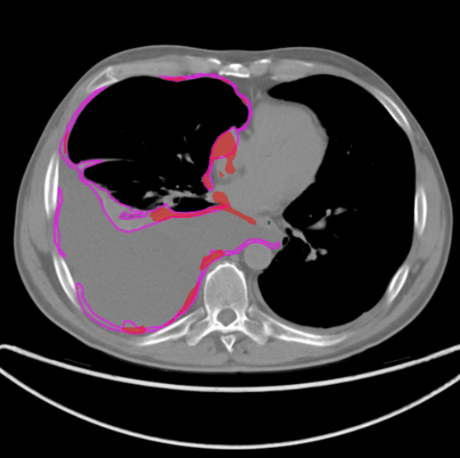} 
        \caption{} 
        \label{fig:TopLeft}
    \end{subfigure}
    \begin{subfigure}{.4\textwidth}
        \includegraphics[width=\linewidth]{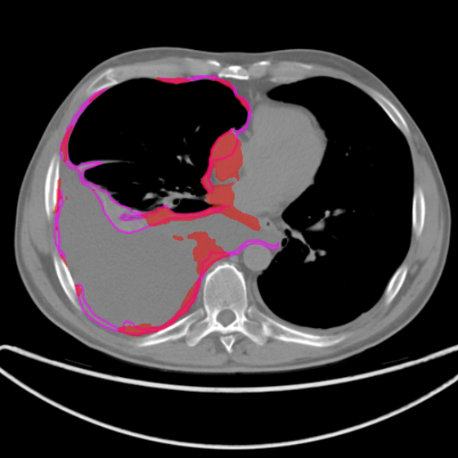} 
        \caption{} 
        \label{fig:TopRight}
    \end{subfigure}
    \caption{Differing contours on the same section of the same patient created with an adjustment in the CNN probability threshold. Purple represents the radiologist’s reference outline, and red represents the CNN pixel-wise segmentation prediction of tumor with in (\subref{fig:TopLeft}) a probability threshold of 0.5 (average DSC: 0.357) and (\subref{fig:TopRight}) a probability threshold of 0.001 (average DSC: 0.476). 
    }
    \label{fig:probmaps}
\end{figure}

\begin{figure} [H]
   \begin{center}
   \begin{tabular}{c} 
   \includegraphics[height=7.5cm]{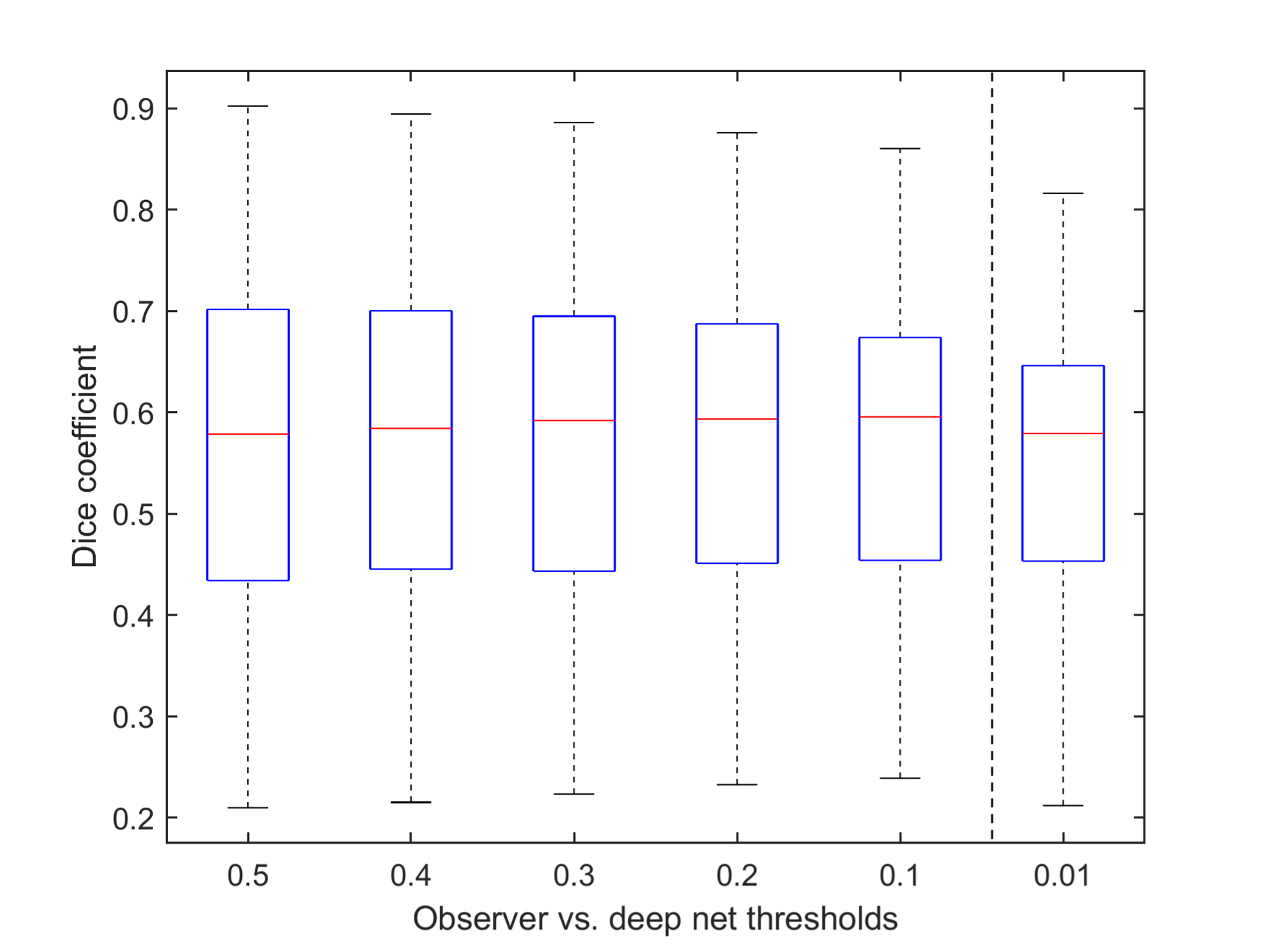}
   \end{tabular}
   \end{center}
   \caption{Boxplots showing the DSC values obtained for tumor comparisons acquired between the radiologist and the deep CNN at six different thresholds. The red lines display the median of the data. }
   \label{fig:Dice full}
\end{figure}

Figure \ref{fig:probmaps} displays a visual representation of a change in the probability thresholds and its impact on the tumor contour. Pleural effusions present were difficult for the CNN to fully appreciate as shown. Overall, the thresholds ranged from 0.001 to 0.9; however, the CNN never assigned a pixel a probability of 0.75 or greater. Figure \ref{fig:Dice full} shows boxplots of the DSC values comparing the reference contours to six of the thresholds applied to the CNN generated probability maps. Except for the 0.01 threshold, the range of DSC values decreased with the incremental reduction of probability thresholds. The median does not substantially change (see Table \ref{tab:full datainfo}).  

\begin{figure} [H]
   \begin{center}
   \begin{tabular}{c} 
   \includegraphics[width=\textwidth]{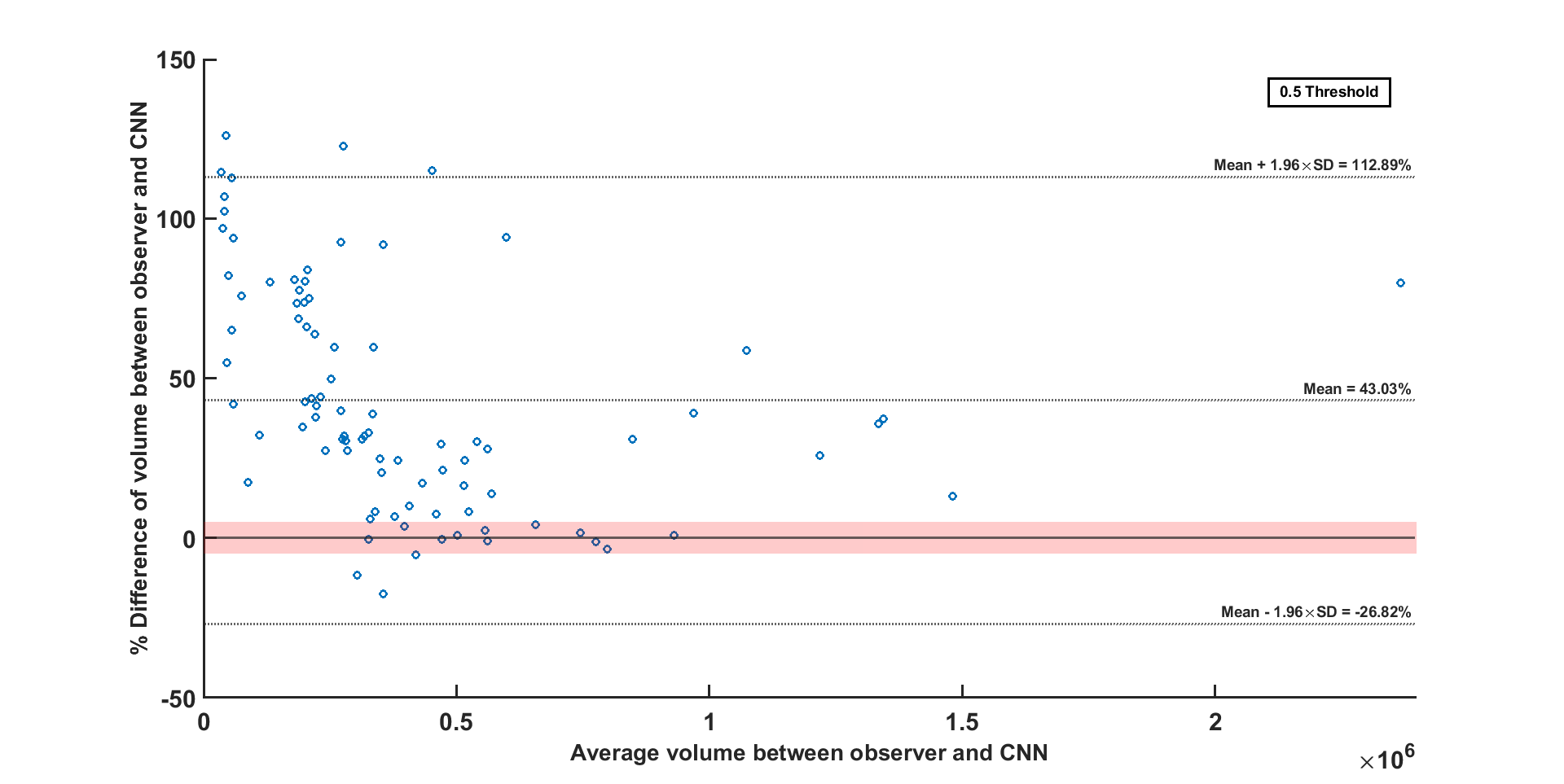}
   \end{tabular}
   \end{center}
   \caption{Bland-Altman plot of the relative differences between manual and CNN-based tumor volumes. The red band highlights differences within 5\% of 0.}
   \label{fig:BA plot}
\end{figure} 

Figure \ref{fig:BA plot} is a Bland-Altman plot \cite{BlandAltman} at the 0.5 threshold displaying the relative percent difference of volume as calculated using the radiologist’s and the CNN contours. The CNN consistently undermeasured the tumor at that threshold, which resulted in a mean percent difference of 43\% (range from -17\% to 126\%, median of 34\%). Eleven (12.5\%) scans were within $\pm$5\% of 0\% difference as shown by the red band in Figure \ref{fig:BA plot}. 

\begin{table}[H]
\centering
\caption{Absolute percent difference ($\pm$ standard deviation) of volume, average, and median DSC at six thresholds. (IQR = interquartile range).}
\label{tab:full datainfo}
\begin{adjustbox}{width=1\textwidth}
{\renewcommand{\arraystretch}{2}
\begin{tabular}{l c c c c c c}
\toprule
     & \multicolumn{1}{c}{Threshold = 0.5} & \multicolumn{1}{c}{Threshold = 0.4} & \multicolumn{1}{c}{Threshold = 0.3} & \multicolumn{1}{c}{Threshold = 0.2}  & \multicolumn{1}{c}{Threshold = 0.1} & \multicolumn{1}{c}{Threshold = 0.01} \\ \midrule
Average absolute \% difference of volume & 43.96 $\pm$ 34.48 & 38.05 $\pm$ 31.78 & 32.68 $\pm$ 28.69 & 27.26 $\pm$ 25.38 & 24.18 $\pm$ 18.87 & 28.16 $\pm$ 18.41 \\
Average DSC & 0.58 $\pm$ 0.17 & 0.58 $\pm$ 0.17 & 0.58 $\pm$ 0.16 & 0.58 $\pm$ 0.15 & 0.58 $\pm$ 0.15 & 0.54 $\pm$ 0.14 \\
Median DSC (IQR) & 0.58 (0.27) & 0.58 (0.26) & 0.59 (0.25) & 0.59 (0.24) & 0.60 (0.22) & 0.58 (0.19) \\ \bottomrule
\end{tabular}}
\end{adjustbox}
\end{table}
Table \ref{tab:full datainfo} shows that the average absolute percent difference of volume consistently decreased with a lowered threshold, further demonstrating the CNN’s undermeasurement of the tumor at the 0.5 threshold. This trend did not occur at the 0.01 threshold, with comparable DSCs but larger percent difference. The average DSC peaked at the 0.3 threshold, while the median DSC reached its maximum at 0.1. Figure \ref{fig:matrices} below displays all the relevant p-value comparisons performed for the percent difference of volume and DSC values. Absolute percent difference of volume was affected more by changes in threshold as compared with DSC. 

\begin{figure}[H]
    \centering
    \begin{subfigure}[t]{0.5\textwidth}
        \centering        
        \includegraphics[height=3.5in]{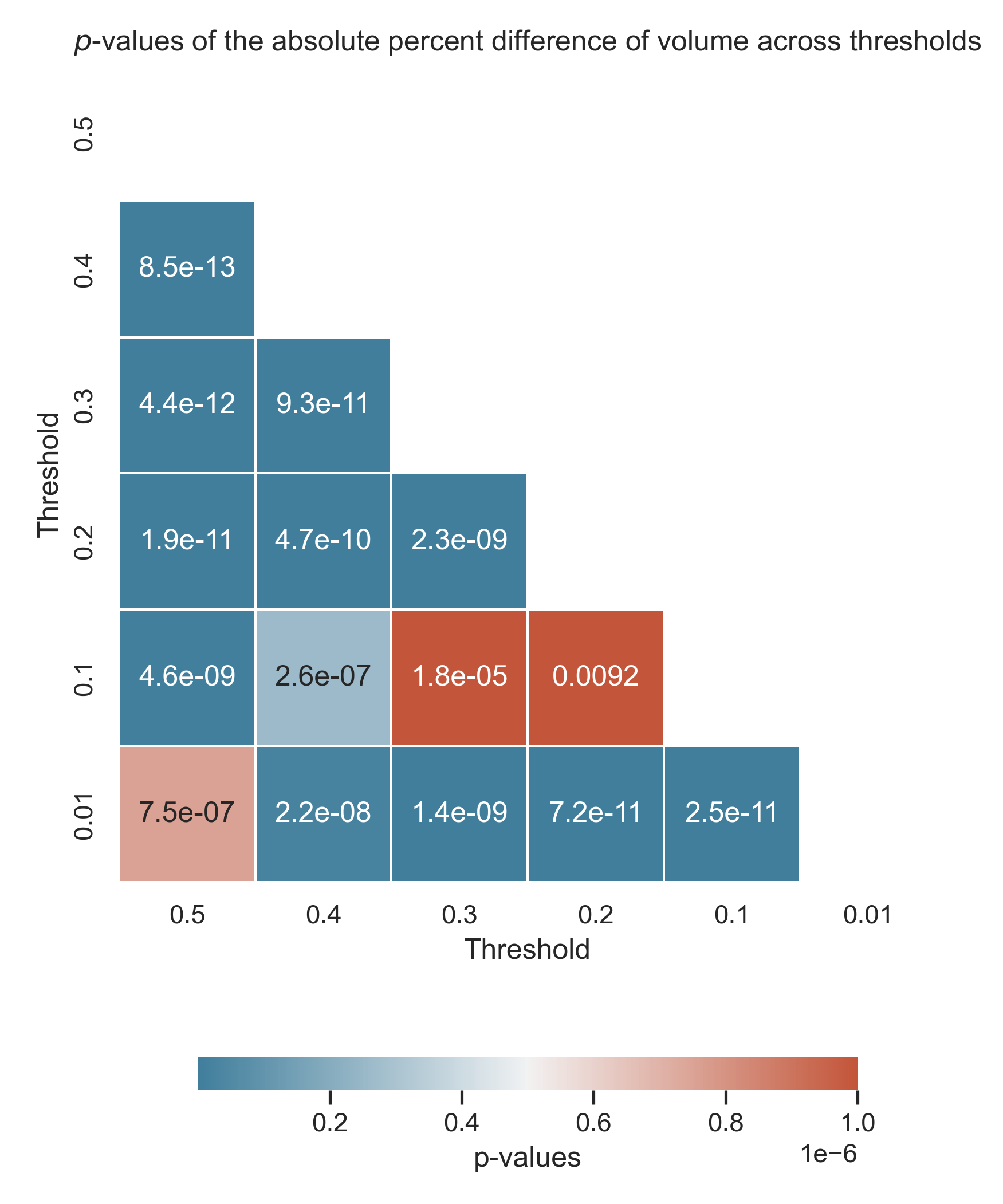}
        \caption{}
        \label{subfig:p-value volume}
    \end{subfigure}%
    ~ 
    \begin{subfigure}[t]{0.5\textwidth}
        \centering        
        \includegraphics[height=3.5in]{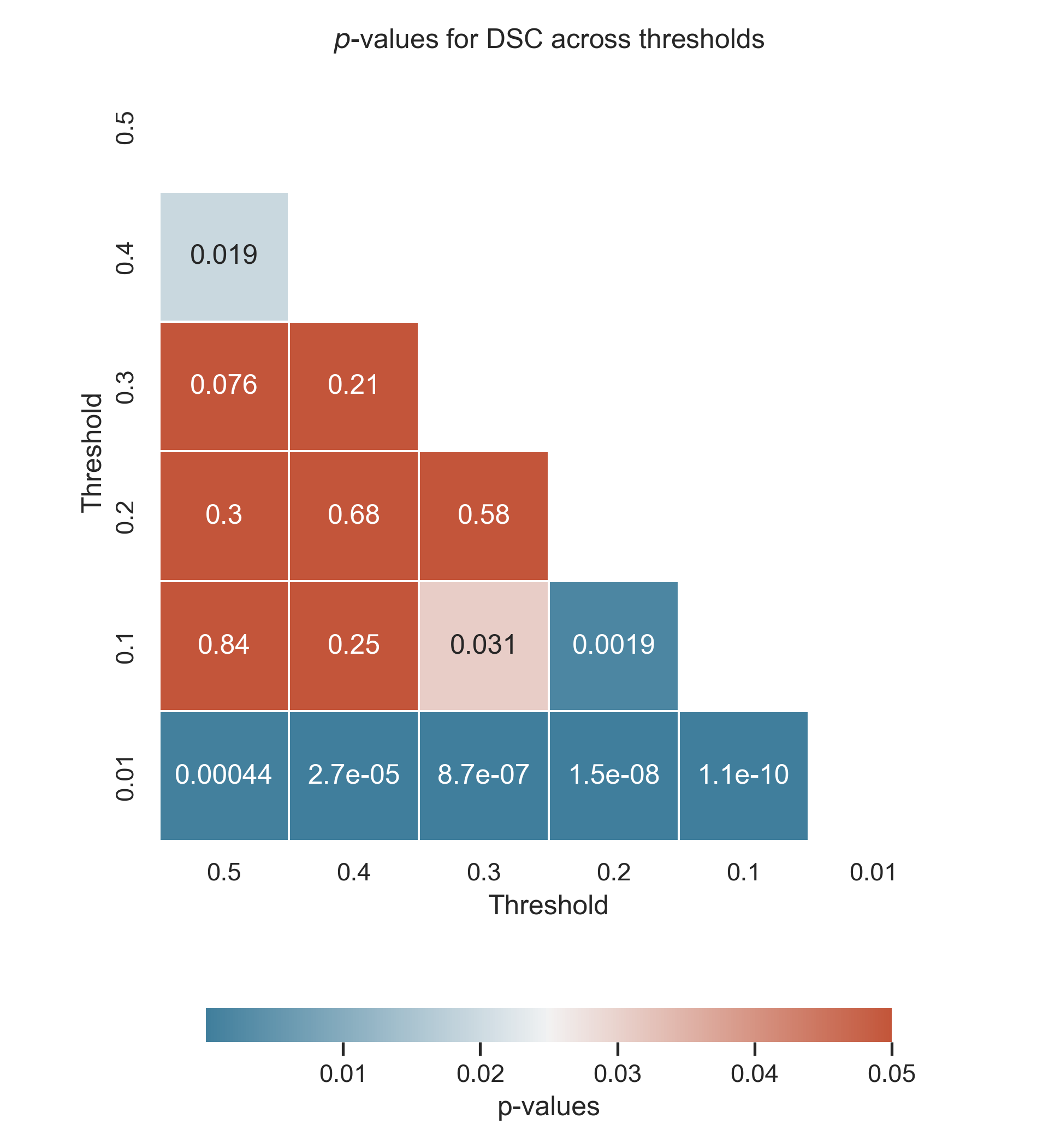}
        \caption{}
        \label{subfig:p-value Dice}
    \end{subfigure}
    \caption{Matrix displaying the p-values when comparing the absolute percent difference of volume (\subref{subfig:p-value volume}) and DSC (\subref{subfig:p-value Dice}) across thresholds.}
    \label{fig:matrices}
\end{figure}
\begin{figure} [H]
   \begin{center}
   \begin{tabular}{c}
   \includegraphics[width=\columnwidth]{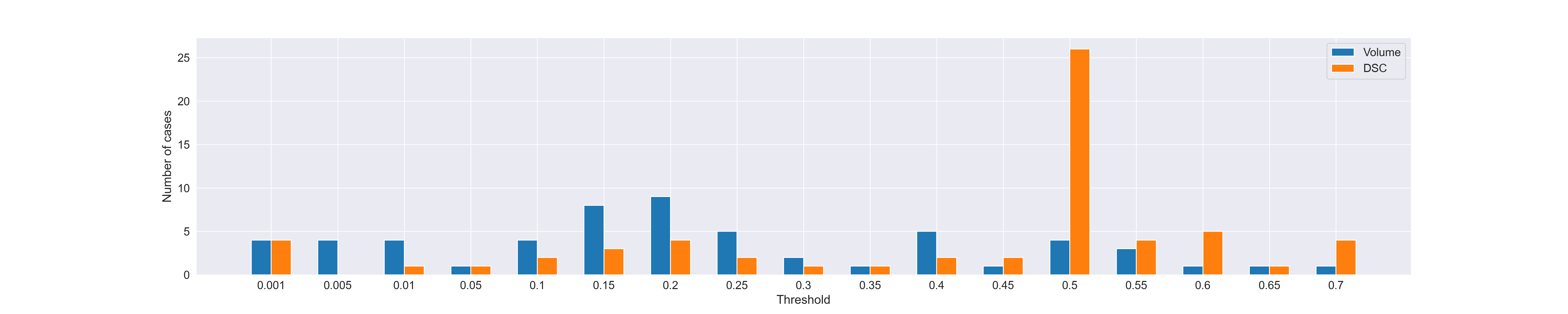}
   \end{tabular}
   \end{center}
   \caption{Histogram of the CNN output thresholds that ``optimize'' for DSC and percent difference of volume.}
   \label{fig:histogram}
\end{figure}

 
Figure \ref{fig:histogram} shows the frequency of a threshold being chosen as ``optimal,'' ranging from 0.001 to 0.7. An optimal threshold differed based on the metric. For example, a certain threshold being optimal for volume meant this was the threshold at which we acquired the lowest percent difference of tumor volume between the radiologist’s and CNN contours; similarly, an optimal threshold for DCS indicated that it maximized DSC. Beside the substantial peak present at a threshold of 0.5 for the DSC, there does not appear to be a distinct pattern of ``best'' values.

\begin{figure} [h!]
   \begin{center}
   \begin{tabular}{c}
   \includegraphics[height=12cm]{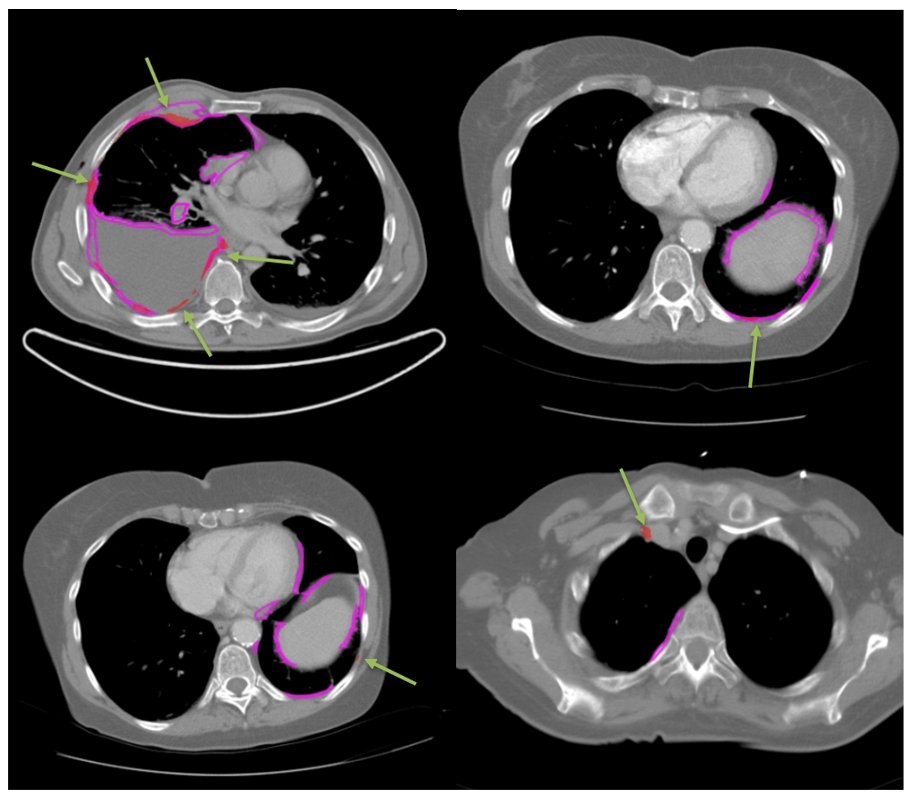}
   \end{tabular}
   \end{center}
   \caption{Example images of the four cases that were greater than the 95\% agreement limits as shown in the Bland-Altman plot (Figure \ref{fig:BA plot}). Green arrows point to regions where the CNN predicted tumor at the 0.5 threshold. Purple outlines are the radiologist’s reference contours.}
   \label{fig:combinedpics}
\end{figure} 

Figure \ref{fig:combinedpics} displays example images from four scans (three patients) that were greater than the 95\% agreement limits as presented in the Bland-Altman plot in Figure \ref{fig:BA plot}. The CNN used was trained on cases that were applicable to mRECIST measurements and was not designed to consider tumors present near the diaphragm, near the lung apices, or invading other parts of thoracic anatomy. Therefore, it unreliably identified tumor for cases with severe pleural effusion as well as disease superior to the aortic arch and inferior to the pulmonary vein, where other anatomic structures complicate the morphology of MPM. 

\begin{figure} [h!]
   \begin{center}
   \begin{tabular}{c} 
   \includegraphics[height=7.5cm]{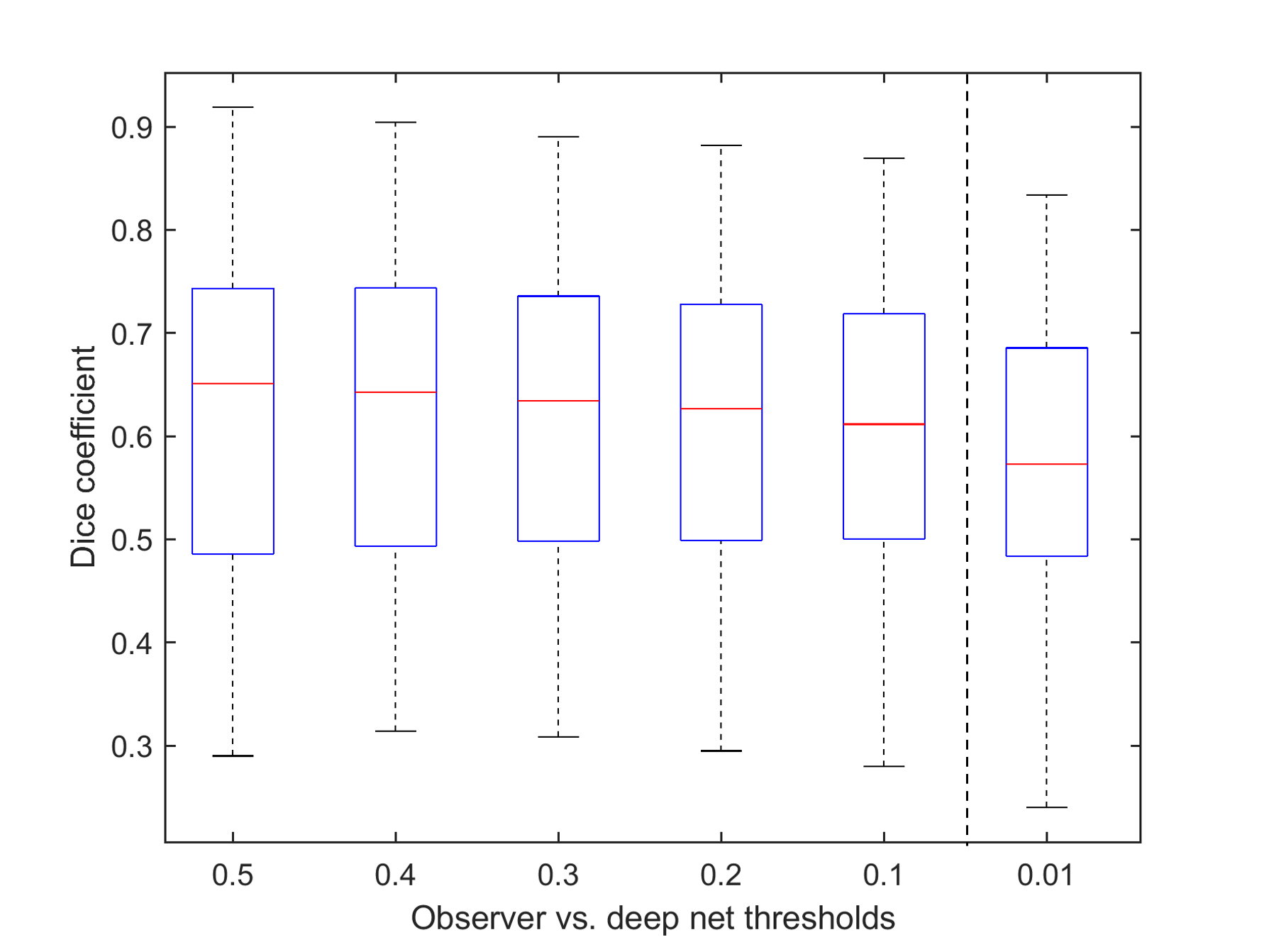}
   \end{tabular}
   \end{center}
   \caption{Boxplots showing the DSC values obtained for the subset tumor comparisons acquired between the radiologist and the deep CNN at six different thresholds. Red lines display the median of the data.}
   \label{fig:Dice subset}
\end{figure}

To exclude such regions, we performed the volume and DSC comparisons only considering CT sections inferior to the aortic arch and superior to the pulmonary vein. Figure \ref{fig:Dice subset} and Table \ref{tab:subset datainfo} parallel Figure \ref{fig:Dice full} and Table \ref{tab:full datainfo}, showing DSC and volume values for the subset analysis. Figure \ref{fig:Dice subset} displays the DSC across the same six thresholds analyzed, and Table \ref{tab:subset datainfo} displays the average absolute percent difference of volume along with average and median DSC values across the thresholds. The range of DSC values is also smaller for the subset analysis (range from 0.24 to 0.92 for the subset as opposed to 0.20 to 0.90).

\begin{table}[H]
\centering
\caption{Absolute percent difference ($\pm$ standard deviation) of volume, average, and median DSC at six thresholds for the subset analysis. (IQR = interquartile range).}
\label{tab:subset datainfo}
\begin{adjustbox}{width=1\textwidth}
{\renewcommand{\arraystretch}{2}
\begin{tabular}{l c c c c c c}
\toprule
     & \multicolumn{1}{c}{Threshold = 0.5} & \multicolumn{1}{c}{Threshold = 0.4} & \multicolumn{1}{c}{Threshold = 0.3} & \multicolumn{1}{c}{Threshold = 0.2}  & \multicolumn{1}{c}{Threshold = 0.1} & \multicolumn{1}{c}{Threshold = 0.01} \\ \midrule
Average absolute \% difference of volume & 44.00 $\pm$ 31.10 & 37.25 $\pm$ 29.10 & 30.81 $\pm$ 27.12 & 25.35 $\pm$ 24.08 & 22.18 $\pm$ 19.26 & 30.55 $\pm$ 19.39 \\
Average DSC & 0.62 $\pm$ 0.18 & 0.62 $\pm$ 0.17 & 0.62 $\pm$ 0.16 & 0.62 $\pm$ 0.15 & 0.61 $\pm$ 0.15 & 0.57 $\pm$ 0.14 \\
Median DSC (IQR) & 0.65 (0.26) & 0.64 (0.25) & 0.63 (0.24) & 0.63 (0.23) & 0.61 (0.22) & 0.57 (0.20) \\ \bottomrule
\end{tabular}}
\end{adjustbox}
\end{table}

Overall, the subset analysis displayed smaller volume percent differences and narrower standard deviations when compared with values in Table \ref{tab:full datainfo}. Further, the average and median DSCs were consistently larger than those acquired when evaluating the entirety of the scan. The percent volume difference and median DSC value at the 0.01 threshold, however, performed worse for the subset. For instance, the percent difference of 30.55$\pm$19.39\% was larger than 28.16$\pm$18.41\% and the median DSC of 0.57 (0.20) was lower than 0.58 (0.19). Though, the average DSC of 0.57$\pm$0.14 for the subset was larger than 0.54$\pm$0.14 when considering the entirety of the scan at the 0.01 threshold.

\section{Discussion}
At lower probability thresholds, more pixels were counted as tumor within a CT section, as demonstrated in Figure \ref{fig:probmaps}. As a result of a lowered threshold, the computed tumor volume increased (Equation \ref{eq:volume}). Furthermore, except for eight cases, the radiologist’s volumes were consistently larger than those of the CNN (Figure \ref{fig:BA plot}), indicating that the CNN did not capture the tumor to its true extent at the default threshold of 0.5. Lastly, as presented in Table \ref{tab:full datainfo}, the absolute percent difference in volume significantly decreased (Figure \ref{subfig:p-value volume}) with a lowered threshold, indicating an increase of calculated tumor volume.

While the absolute percent difference decreased at smaller thresholds, the average and median DSC were not different, both within 0.58-0.60. These values are slightly higher than in the current literature \cite{Kidd}. Therefore, the accuracy of contour overlap did not suffer with the change of thresholds. In other words, the newly included pixels at lower thresholds were not at arbitrary locations within the scan but were, on average, at relevant anatomic regions that overlapped the reference contours, which demonstrated the robustness of the CNN in identifying tumor, even for lower pixel probabilities. A limitation in this analysis was the inherent bias present for the DSC calculation: rather than delineating the tumor on the original CT scans, the radiologist modified the contours generated by the CNN at a threshold of 0.5, which explains the distinct peak at the 0.5 threshold when seeking to maximize DSC (Figure \ref{fig:histogram}). 

We obtained large 95\% limits of agreement due to the considerable standard deviation of percent differences of volume. Four cases exceeded the 95\% limits of agreement (Figure \ref{fig:combinedpics}). The CNN failed to capture disease that surrounded organs such as the spleen, vertebral column, and heart in those scans. For these cases, the radiologist provided new contours to capture regions of tumor excluded by the CNN at the 0.5 threshold, to establish a reference standard, instead of modifying a preexisting CNN output. Severe pleural effusion was erroneously identified as tumor. Overall, these four examples underly major trends of where the model underperformed. Upon closer examination of the images, the CNN had difficulty contouring disease for the aforementioned reasons along with presence of metallic artifacts, disease in the fissure, and disease present superior to the aortic arch and inferior to the pulmonary veins. 

Poor performance was expected for regions that were at the level of the diaphragm or superior to the lung apices because the model was not trained using such regions. Rather, the model was evaluated using tumor contours applicable to mRECIST measurements. To account for this discrepancy, we performed a subset analysis on sections only between the aortic arch and superior to the pulmonary veins. Beside the 0.5 and 0.01 threshold, the percent differences of volume were smaller, and the DSCs were higher across the six thresholds evaluated (except for the 0.01 threshold where the entire scan’s median DSC was slightly higher). Further, the median DSC was highest at the 0.5 threshold and decreased with lower thresholds. This may be due to the training of the model once more, as it was trained using the 0.5 threshold. Therefore, we acquired performance comparable to what the model was trained on previously. It is also important to note that the 0.01 threshold acquired the lowest DSC and increased the percent volume difference for the whole scan along with the subset analysis. This indicated that the delineations at low thresholds erroneously considered a substantial number of pixels as tumor, pixels that encroached on anatomy in the region. Overall, for both the subset and whole scan, the percent difference of volume, average DSCs, and median DSCs decreased with a decrease in threshold, with the subset analysis outperforming the entire scan for both metrics. 

There are some limitations that should be addressed in this work. First, as previously mentioned, by displaying for the radiologist contours that had been generated at the 0.5 threshold, there was an inherent bias: this bias has been shown to impact the modified outlines produced by observers \cite{Sensakovic}. The 0.5 threshold was chosen a priori as it is often selected because it is an intuitive value to binarize output probability maps. The aim of this study was to determine the impact of the probability threshold on the two figures of merit studied, not necessarily to study the clinical implementation or the generalizability of a given threshold; therefore, this study was not hindered by having a reference standard with only a single radiologist. A second limitation was the performance of the segmentation task using a 2D CNN architecture as opposed to more advanced techniques, e.g., a 3D architecture. This is will be explored in future work, as the current dataset size may restrict model complexity (compared with a 2D architecture) and result in poor performance when implementing a 3D architecture.

Overall, the purpose of this work was to study the impact of CNN probability map thresholds on the percent volume differences and DSCs when comparing CNN-generated mesothelioma tumor outlines with a radiologist’s tumor outlines. These results indicate that the investigation is slightly more nuanced, as lowering the probability threshold (1) predictably increased the resulting tumor volume (which was consistently smaller than the radiologist’s) and lowered the percent difference but (2) only negligibly affected the DSC. It is important to make the distinction between CNN volume measurement and DSC. Ensuring that the CNN acquires volume comparable to the reference standard is critical, as volume can be used to capture tumor burden more completely for response assessment. However, we must also be cognizant of the spatial regions where the CNN identifies tumor; it is not sufficient only to match volumes with the reference standard, but also to match the location of the contours. Future work will train the CNN using cases for which this current model was deficient, providing it with cases that displayed pleural effusion and disease surrounding the various structures in the thorax. 

\section{Conclusion}
This study explored the impact of changing the threshold applied to the probability maps output by a CNN when segmenting MPM tumors on CT scans. After investigating thresholds from 0.001 to 0.9, a clear peak at the 0.5 threshold was found for DSC; however, there was no definitive threshold value to minimize the percent difference of volume between the radiologist’s and CNN outlines. The percent differences of volume decreased when lowering the probability threshold, while the median DSC did not significantly change. The CNN performance was deficient on scans that contained severe pleural effusion and disease that bordered other structures in the thorax. Therefore, a subset analysis was conducted, which yielded improved results for difference of volume and DSC. Overall, this pilot study highlighted the impact of varying CNN-generated probability map thresholds on mesothelioma tumor outlines, using percent difference volume and the DSC as the figures of merit.

\section*{Acknowledgments}
Support: Research reported in this publication was supported by the National Cancer Institute of the National Institutes of Health under Award Numbers U10CA180821 and U10CA180882 (to the Alliance for Clinical Trials in Oncology), UG1CA189863, UG1CA2333320, and UG1CA233253. https://acknowledgements.alliancefound.org. The content is solely the responsibility of the authors and does not necessarily represent the official views of the National Institutes of Health. Also supported, in part, by John D. Cooney and the firm of Cooney and Conway through the University of Chicago Comprehensive Cancer Center.


\bibliographystyle{unsrt}  
\bibliography{references}

\end{document}